# Programmable black phosphorus image sensor for broadband optoelectronic edge computing

Seokhyeong Lee[1,#], Ruoming Peng[1,#,*], Changming Wu[1], and Mo Li[1,2,*]

[1]Department of Electrical and Computer Engineering, University of Washington, Seattle, WA 98195, USA

[2]Department of Physics, University of Washington, Seattle, WA 98195, USA

**Abstract**

Image sensors with internal computing capability enable in-sensor computing that can significantly reduce the communication latency and power consumption for machine vision in distributed systems and robotics. Two-dimensional semiconductors are uniquely advantageous in realizing such intelligent visionary sensors because of their tunable electrical and optical properties and amenability for heterogeneous integration. Here, we report a multifunctional infrared image sensor based on an array of black phosphorous programmable phototransistors (bP-PPT). By controlling the stored charges in the gate dielectric layers electrically and optically, the bP-PPT's electrical conductance and photoresponsivity can be locally or remotely programmed with high precision to implement an in-sensor convolutional neural network (CNN). The sensor array can receive optical images transmitted over a broad spectral range in the infrared and perform inference computation to process and recognize the images with 92% accuracy. The demonstrated multispectral infrared imaging and in-sensor computing with the black phosphorous optoelectronic sensor array can be scaled up to build a more complex visionary neural network, which will find many promising applications for distributed and remote multispectral sensing.

* Corresponding author: pruoming@uw.edu, moli96@uw.edu

#These authors contributed equally to this work.

**Introduction**

2D semiconductors have tremendous potential in optoelectronics because they afford a wide range of bandgaps with tunable optoelectronic properties[1–3]. Being atomically thin and transferable, they are amenable to heterogeneous integration with photonic circuits and microelectronics to realize advanced functionalities[4–9]. In terms of broadband operation in the infrared, black phosphorus (bP) stands out for its tunable bandgap corresponding to a wide infrared spectral range. Discrete, array, and waveguide-integrated bP photodetectors with compelling performance have been demonstrated for the infrared[10–17]. Leveraging its broadband infrared responses, arrays of bP photodetectors can be utilized for multispectral imaging, which acquires spatial images with simultaneous spectral information[17-18]. Multispectral imaging combined with artificial neural networks (ANN) has become a powerful tool for biomedical imaging[19–22], fresh food classification[23–25], and surface damage detection in industrial sites[26–28]. This imaging technique can generate a tremendous amount of data and is consequently computation-intensive and latency-sensitive, and thus can benefit from the emerging scheme of edge computing[9,29–32]. Preprocessing the images within the sensors at the edge rather than in the cloud can largely alleviate the data streaming load to the servers, improving the bandwidth budget[33–35] and reducing latency and power consumption. These advantages of edge computing have urged the development of novel optoelectronic edge sensors that combine vision-sensory and computational functionalities in the same devices[8,9], which recently have been demonstrated using 2D transition metal dichalcogenides (TMDs) for visible spectral imaging. Realizing such a scheme using bP will extend it to the infrared spectral range, enabling intelligent night vision and multispectral sensing.

Here, we present a multifunctional image sensor that combines the functions of multispectral imaging and analog in-memory computing to implement an in-sensor ANN for image recognition. The image sensor is based on an array of programmable phototransistors made of few-layer bP (bP-PPTs), which are sensitive to a broad infrared spectral range from 1.5 to 3.1 μm in wavelength. The bP-PPT's programmability and memory stem from the stored charges in the rationally designed stack of gate dielectrics that have a long retention time and effectively modulate the photoconductance of the bP channel. The sensor can be programmed and read out both electrically and optically, enabling in-memory computing and remote programming. It is used as an optical frontend that can capture multispectral images in the infrared and perform image processing and classification tasks (Fig. 1a), demonstrating its promise for distributed and remote sensing applications at household, farming, and industrial sites.

**Result**

Fig. 1b depicts the structure of a single bP-PPT device, which consists of a few-layer bP flake as the channel, a stack of $Al_2O_3/HfO_2/Al_2O_3$ (AHA) as the gate dielectric and charge storage layer[36–38], and a top gate electrode. Contrary to the conventional floating gate devices that trap charges on an isolated metallic gate, the bP-PPT stores charges in the $HfO_2$ dielectric layer with a lower barrier height, which offers more reliable and faster operation and simplifies the fabrication process [38–41]. For optical access to the bP-PPT, indium-tin-oxide (ITO) is used as the transparent top gate electrode. Fig. 1d depicts the band alignment of the multiple layers in the bP-PPT device. We design the layer structure and select the materials such that charges (electrons or holes) can tunnel from the top gate through the thin $Al_2O_3$ barrier layer to be stored in the $HfO_2$ layer and effectively modulate the bP channel with field effect. Moreover, the electron affinity difference between $Al_2O_3$ and $HfO_2$ ($\chi_{Al2O3}$= 1.0 eV, $\chi_{HfO2}$= 2.5 eV) determines the storing energy to be ~1.5 eV, enabling optical control of the stored charges—they can be removed by illuminating with visible light ($\lambda$ < 0.886 μm) but not with infrared light in the telecommunication band or longer wavelength. The conductance and photoresponsivity of the bP-PPT are controlled by the density of the trapped charges[12–14], thus can be set either by applying electrical gate voltage or by shining visible spectral light pulses, enabling local and remote programming of the device.

To construct an image sensor and processor, we fabricated a 4 × 3 array of bP-PPTs on a single bP flake, as shown in Fig. 1c (see Methods for more details about the fabrication process). We optimized the thickness and the deposition process of the AHA multilayer to realize a high charge storage density of $1.8×10^{13}$ cm$^{-2}$ (see Supplementary note 2 for the method used to determine the density). Fig. 1e shows the collective measurement results of the source-drain current ($I_{ds}$) of the devices in the array when the gate voltage ($V_G$) is swept. Because charges are injected and stored in the AHA multilayer during the $V_G$ sweep, the $I_{ds}$-$V_G$ curve shows a hysteresis loop with a large memory window of 25 V in $V_G$. The high charge storage density leads to effective control in the electrical conductance of the bP-PPT, achieving an on/off ratio > 200[36]. Since the array is fabricated on the same bP flake, it has excellent uniformity that inter-device variation in the on/off ratio among 9 devices is less than 8% (inset, Fig. 1e).

Fig. 2a illustrates the working principle of electrically programming the bP-PTTs by applying voltage pulses to the gate. The device can first be reset with a depressive pulse (-18 V amplitude, 50 ms duration) to a fully-off state (state #0) with low conductance. Afterward, it can be programmed by applying positive voltage pulses with amplitude in the range of 10 ~ 18 V and a fixed duration of 20 ms. By varying the pulse amplitude, the device can be programmed to states of more than 8 distinguishable levels (equivalent to 3 bit) in its conductance when the bP channel is changed from n-type to p-type doping (Fig. S1, Supplementary Information). Fig. 2c and d show results of four representative states with a long retention time >4000 s (Fig. 2c) and linear *I-V* characteristics (Fig. 2d). The latter is important for its application in analog computing.

Even higher precision can be achieved by programming the devices optically because optical pulses can directly excite the stored charges to remove them, and the duration of optical pulses can be controlled more accurately than voltage pulses. We demonstrate optical programming of the bP-PPT devices using optical pulses in wavelength of 780 nm, which provides sufficient energy to activate the stored charges to overcome the trapping potential (Fig. 1d). Before programming, the bP-PPT is initialized electrically to state #0. Subsequently, it is illuminated with optical pulses with fixed average power (~10 µW at the device) and varying duration so the pulse energy is varied between 10 nJ and 2 µJ. As shown in Fig. 2e, these optical pulses program the bP-PPT to 36 states with different levels in conductance to represent 5 digital bits, a record-high number of levels achieved in charge storage devices. The programming process is accurate, arbitrary, and repeatable. The inset of Fig. 2e shows three adjacent levels that can be programmed repeatedly with high precision. Furthermore, each programmed state is stable for >1000 s, sufficient for analog computation applications (see Supplementary Information).

The narrow bandgap of bP enables the bP-PPTs to be operated as broadband photodetectors that can detect optical signals from the near-infrared (NIR) to the mid-infrared (MIR) spectral range. Earlier studies have reported that a bP photodetector's responsivity is sensitive to the doping level and type of the bP channel[11–14]. In our bP-PPTs, since we can control the density of the stored charge to modulate the doping level and type of the bP channel, we can program their photoresponsivity in the same way as their electrical conductance. Fig. 2f shows the photoresponsivity of a bP-PPT measured in the wavelength range from 1.5-3.1 µm when the device is set to high and low conductance states (corresponding to states #0 and #35 in Fig. 2e), respectively. Note that the low conductance state (state #0) has a high photoresponsivity due to the Burstein-Moss effect[11,14]. The unmeasured spectral range (1.8-2.6 μm) is due to the tunability gap of the light source (M-Square Firefly IR). The linearity of the devices' photoresponse is also verified for an incident optical power of up to 30 mW (Fig. S3 in Supplementary Information). Therefore, the bP-PPT has a programmable photoresponse in all the telecommunication bands (S, C, L bands) and the mid-infrared range.

The above results show that the bP-PPT devices can be programmed both electrically and optically. The programmed state is non-volatile and can be read out either electrically by measuring the device's conductance or optically by measuring its photoresponsivity. In both cases, the device is operated in the linear regime and thus can be utilized for analog computing. Such a hybrid of multifunctional operation modes enables the utilization of a bP-PPT array to implement a mixed-mode optoelectronic neural network system. The same bP-PPT array can act as both the optical frontend to receive and preprocess optical images and an electrical processor with in-memory computing to post-process the images (Fig. 3a).

**bP-PPT in-sensor convolution for edge detection**

We first use the bP-PPT array to detect infrared optical images and preprocess them in the sensor[8,9]. To prove the concept, we configure the bP-PPT array to perform edge detection of images by programming their photoresponsivity ($R$) to represent convolutional kernel matrices and receiving input images transmitted and encoded in the optical power ($P_{in}$) in telecommunication bands. Measuring the photocurrent output $I_{Ph} = R \cdot P_{in}$ from the array corresponds to a multiply-accumulation (MAC) operation[42–44] on the input image with the kernel matrix stored in $R$. For edge detection, the photoresponsivity matrix $R$ is optically programmed to binary values (Fig. 2f) and, after proper normalization, to represent kernel matrix $\begin{bmatrix} -1 & 1 \\ -1 & 1 \end{bmatrix}$ for right edge detection ($\begin{bmatrix} 1 & 1 \\ -1 & -1 \end{bmatrix}$ for top, $\begin{bmatrix} 1 & -1 \\ 1 & -1 \end{bmatrix}$ for left, and $\begin{bmatrix} -1 & -1 \\ 1 & 1 \end{bmatrix}$ for bottom edges) [45–47].

To demonstrate the broadband capability of the bP-PPT array, we encode three different 8-bit grayscale images (Fig. 3b; top: handwritten digits; middle: a husky dog; bottom: a cameraman) using wavelengths in three telecom bands: 1510 nm in the S band, 1550 nm in the C band, and 1590 nm in the L band, respectively. The brightness of each pixel is encoded into the optical power using variable optical attenuators (VOA) and illuminated on a 2×2 bP-PPT array. Each bP-PPT device is set to have a high responsivity of 60 mA/W to represent 1, or a low responsivity of 20 mA/W to represent -1 (Fig. 2f). The measured photocurrents are normalized and offset to calculate the convolution. The convolved images without any further post-processing are shown in Fig. 3c to f, for right, top, left, bottom edges, respectively. Fig. 3g shows the combination of all types of edges, resulting in a clear silhouette of each image. Thus, we demonstrate the bP-PPT array's application as an optical frontend capable of multispectral imaging reception and preprocessing.

**bP-PPT CNN for image recognition**

Besides its photoresponsivity, the conductance of the bP-PPT array can also be programmed to perform MAC operation by measuring the source-drain current $I_{DS} = V_{DS} \cdot g_{bP}$, where $g_{bP}$ is the conductance matrix of the array programmed to represent the weight matrix, $V_{DS}$ is the source-drain voltages applied to the array as the input vector. An optoelectronic convolutional neural network (CNN) thus can be implemented with the array connected to the previous sensory devices, where the optical input image is detected and converted to electrical signals (Fig. 4a, red boundary). In Fig. 2e, we have demonstrated precise programming of the bP-PPT to 36 discrete levels, ensuring high accuracy in weight training and inference calculation[48,49]. Here, we use the 3 × 3 bP-PPT array to demonstrate a CNN that recognizes images of handwriting numbers "0" and "1" from the MNIST data set. The CNN consists of an input layer that captures a 28 × 28-pixel image, a convolution layer with two 3 × 3 kernels, an average pooling layer followed by an 8 × 2 fully connected (FC) layer (Fig. 4a). The network is trained offline with 12,000 images of the training set with 100 epochs, delivering the final output scores that classify the input image to "0" or "1" with 99% accuracy. The trained network model is remotely programmed into the bP-PPT array by illuminating each pixel with the programming optical pulses. The kernel elements are discretized to accommodate the 36 discrete levels of the programmable states and used consistently in the experiment and simulation (Fig. 4b). For example, the element value 2.00 in kernel 1 (K1), the largest element, is represented by setting a pixel of the bP-PPT to state #35 (in Fig. 2e). By optically programming the 9 pixels of the array to the kernel elements, encoding the image pixels in the source-drain voltages, and measuring the source-drain current, the convolution calculation is executed on-chip to obtain the feature maps, followed by the average pooling and FC layers.

The two output nodes from the FC layer are activated with Softmax function and stored as scores to complete the classification task.

To verify the accuracy of the bP-PPT optoelectronic CNN, 100 randomly chosen images of handwritten numbers (48 of "0"s and 52 of "1"s) from the MNIST data set were tested. The results are compared with the simulated results obtained from an offline computer. Note that this simulation is different from the first training with 99% accuracy due to the limited 36-level discreteness of the kernel element values. The bar graph in Fig. 4c compares the experimental and the simulated output scores of the two labels "0" and "1" for 50 test cases, which have shown excellent agreement. The gray-tarnished bars highlight the incorrectly recognized cases. The summary of the experimental and the simulation results are shown as the confusion table in Fig. 4d. The bP-PPT array-based CNN reached a high accuracy of 92%, comparable with the simulated results (95%).

**Conclusion**

To summarize, we have demonstrated a phototransistor array based on bP (bP-PPT) that can be programmed electrically and optically by utilizing the stored charges in the gate dielectric stack that has a long retention time. Particularly, our device has a high programming precision with a resolution higher than 5-bit, which is among the highest of devices based on charged trapping mechanism. Leveraging its flexible functionality, we use the bP-PPT array to realize vision-sensory functions with in-memory computing. The sensors' programmable photoconductivity enables in-sensor computing for edge detection on images that are optically encoded and transmitted over a broad infrared band. The same bP-PPT array can also be electrically programmed on the backend to implement a CNN to perform image recognition tasks. Additionally, the demonstrated programmable photoresponsivity in the near-IR can be extended to a broader range of infrared (currently limited by the laser tunning range).  It will allow multispectral image processing on edge devices, which can expedite many processes in industrial or biomedical applications[20–28]. Furthermore, recently reported centimeter-scale growth of bP suggests that it is promising to scale the bP-PPT array to an even larger array of megapixels[50]. Thus, the demonstrated multifunctional optoelectronic bP-PPT array, combined with parallel imaging and programming schemes, such as spatial light modulation and wavelength division multiplexing, can realize more complex deep neural networks for machine vision sensors distributed with edge computing.

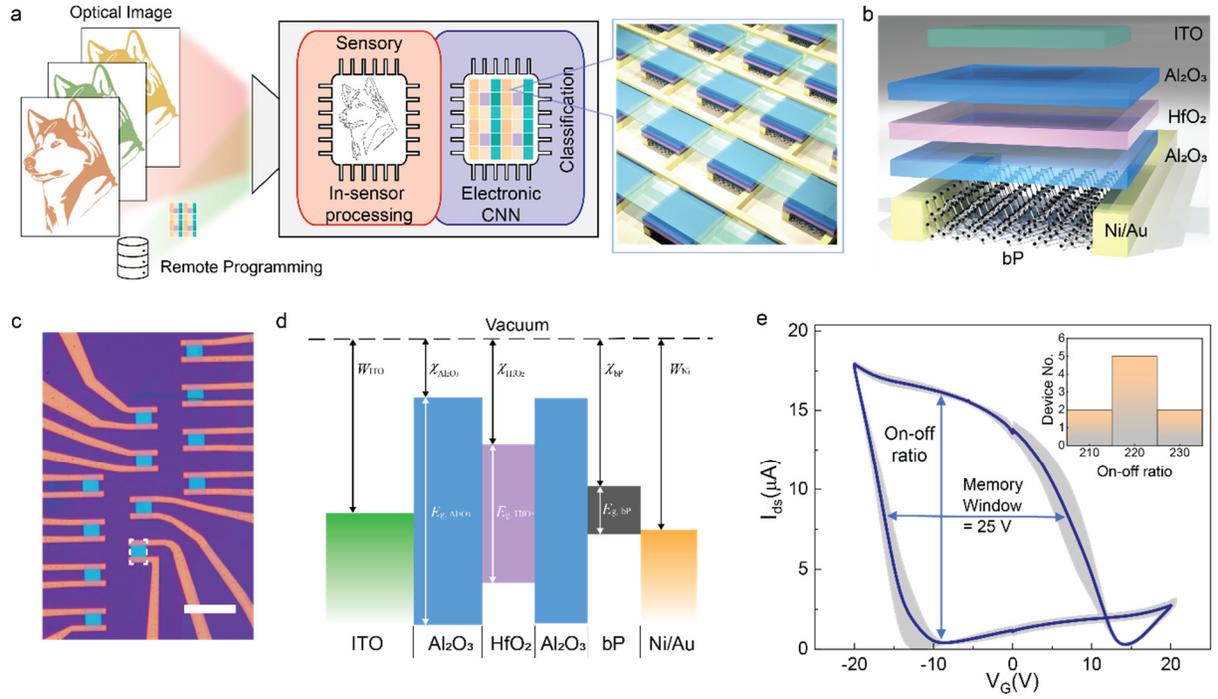

**Figure 1.** Programmable bP phototransistor (bP-PPT) array. **a.** The bP-PPT array is capable of multispectral infrared imaging and is programmable for in-sensor computing. The array can be programmed remotely using optical control signals. **b.** The layer structure of a bP-PPT device. **c.** Optical microscope image of a 3 × 4 array of bP-PPT devices patterned from a single bP flake. White dashed square indicates the patterned bP. Scale bar: 10 $\mu$m. **d.** Band diagram of the bP-PPT device. The AHA dielectric stack is designed to store charges in the $HfO_2$ layer. The electron affinity ($\chi$), bandgap ($E_g$), and work function ($W$) of the layers are: $\chi_{Al_2O_3}$=1.0 eV, $E_{g,Al_2O_3}$=7.7 eV, $\chi_{HfO_2}$= 2.5 eV, $E_{g,HfO_2}$=4.9 eV, $\chi_{bP}$ = 4.4 eV, $E_{g,bP}$ = 0.3 eV, $W_{Ni}$ = 5.01 eV, $W_{ITO}$ = 4.7 eV. **e.** $I_{ds}$-$V_G$ measurement of nine bP-PPT devices. Each bP-PPT shows a large memory window of 25 V in $V_G$ and an on-off ratio of ~200. The inter-device variation (standard deviation) in the array is shown as the gray shaded area. Inset: histogram of the on-off ratio of the devices in the array.

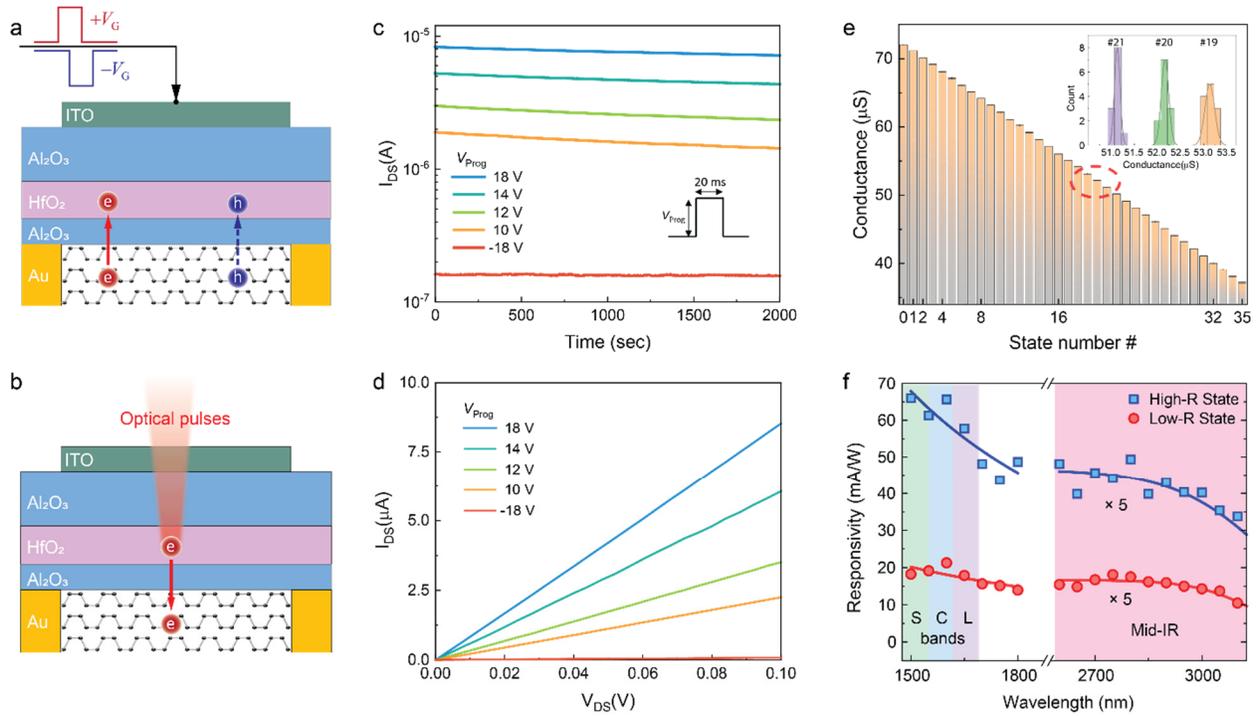

**Figure 2.** Mixed-mode operation of the bP-PPT device. **a, b.** The schematic illustration of the working principle of the programming the bP-PPT with electrical gate voltage pulses (a) and (b) optical pulses. **c.** The bP-PPT can be programmed to 6 states of well-resolved conductance levels using programming pulses of different voltage amplitudes ($V_{prog}$). Note that a negative repressive pulse (-19 V) resets the device to the lowest conductance. **d.** The *I-V* characteristics of the device at each programmed state, showing the linear conductance ($g_{bP}$). **e.** The bP-PPT can be optically programmed using visible pulses to 36 levels in conductance. The inset shows the histogram of the well-separated conductance when the device is programmed repeatedly to three adjacent states. **f.** The bP-PPT's photoresponsivity over the near-IR (including the telecom S,C, and L-bands) and the mid-IR ranges. The discontinued spectral region is due to the gap of the laser tunability. The bP-PPT's photoresponsivity can be programmed to two states when its conductance is set to high (state #0 in **e**) or low (state #35 in **e**).

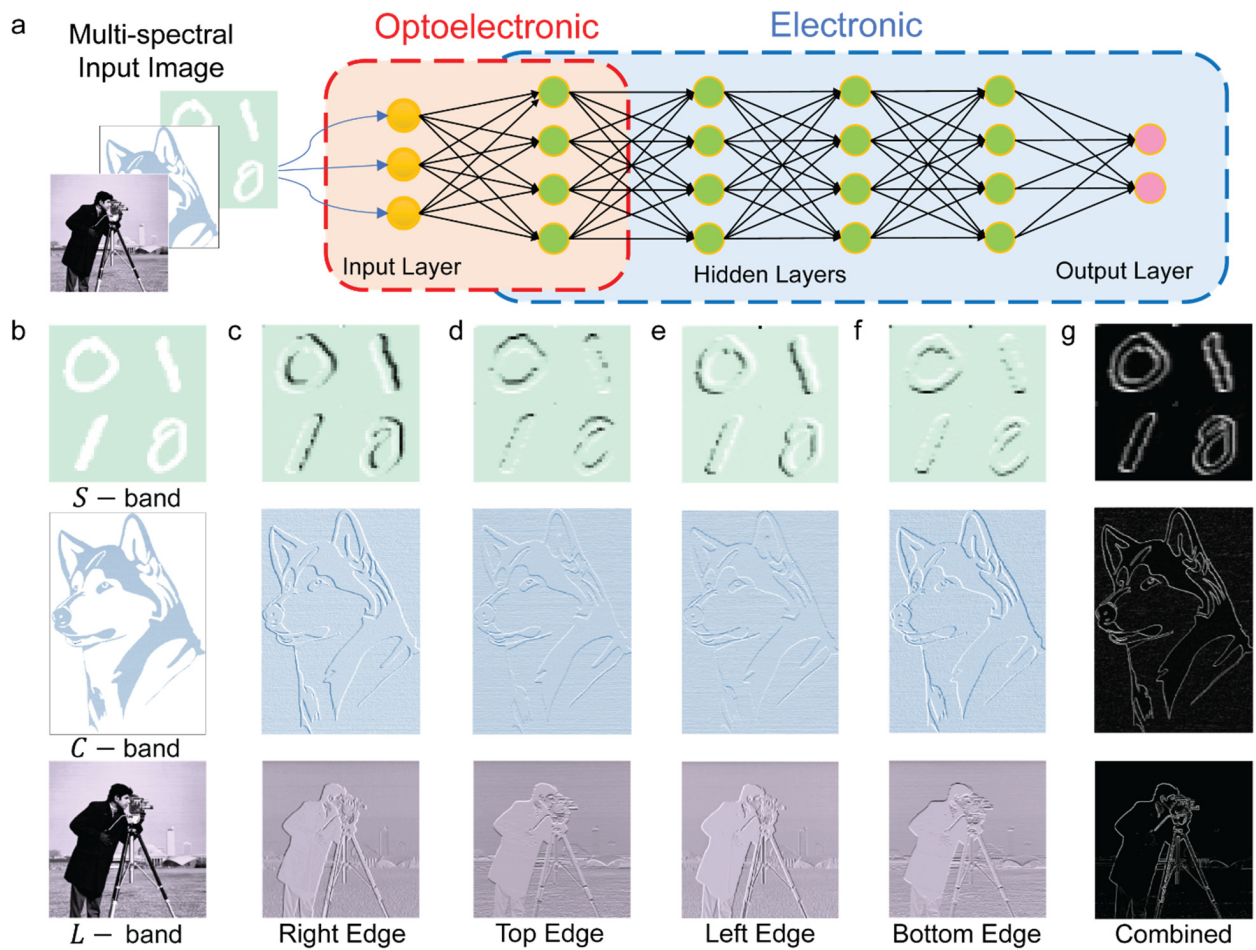

**Figure 3.** bP-PPT array for imaging with in-sensor convolutional kernels. **a.** The bP-PPT array receives imaged in multiple wavelength bands. The array's photoresponsivity matrix is programmed to represent the convolution kernel to directly preprocess the images in the optoelectronic domain (red dashed line box). The array's conductance matrix is then programmed to perform inference computation in the electrical domain (blue dashed line box). **b.** The original input images encoded in the optical power transmitted in three different telecom bands. Top: handwritten digits (56 × 56 pixels, *S*-band); middle: a husky dog (312 × 222 pixels, *C*-band); bottom: a cameraman (256 × 256 pixels, *L*-band). **c-f.** The resultant images after convolution with the right, top, left, and bottom edge kernels, respectively. **g.** The final images combining all the edges.

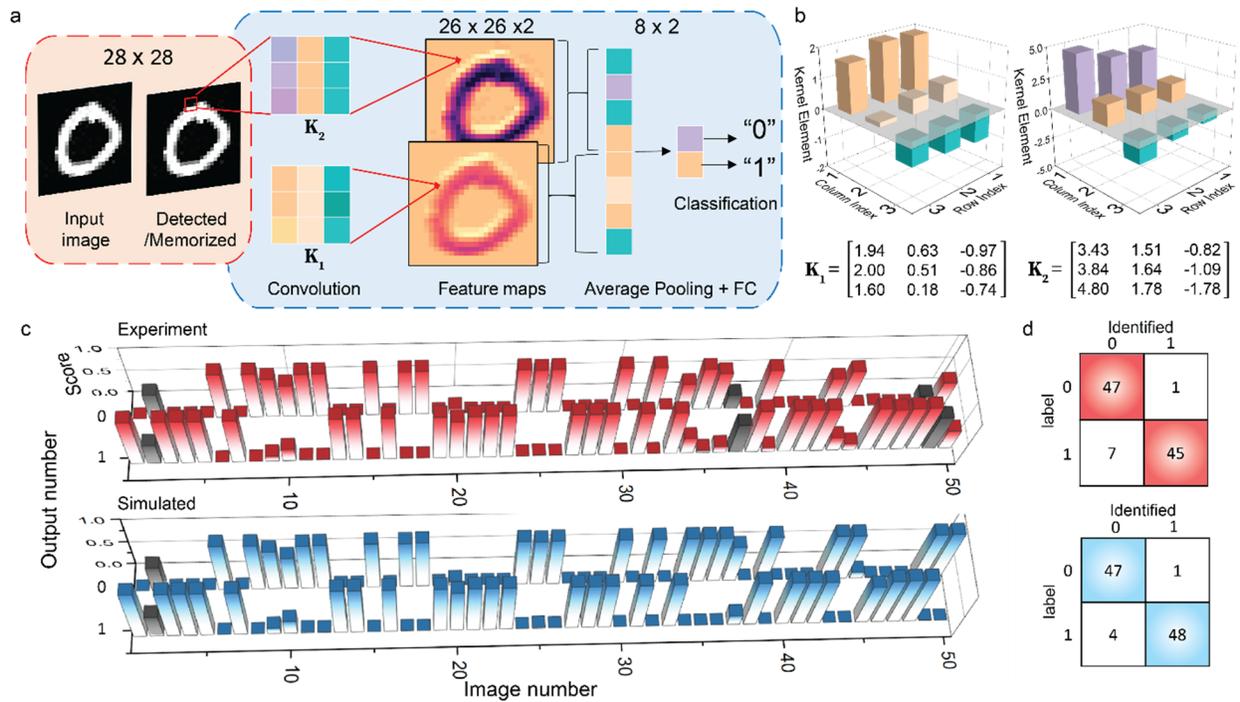

**Figure 4.** bP-PPT array optoelectronic CNN for image recognition. **a.** A CNN model for classifying handwriting numbers "0" and "1" from the MNIST dataset. The CNN consists of two convolution kernels, an average pooling layer, and a fully connected layer. The images captured by the bP-PPT array are further processed by the bP-PPT array in the electrical domain. **b.** The 3×3 bP-PPT array is programmed with 5-bit precision to represent two kernels generated by offline training. **c.** The experimental and simulated results for image recognition using the bP-PPT array. Each bar is the score indicating the possibility of the CNN recognizes an image in the MNIST image library. **d.** The experimental and simulated confusion table for 100 images from the MNIST dataset. Colored diagonal elements in the table indicate the correctly identified cases.